\documentclass[aps,prl,twocolumn,groupedaddress,showpacs,letterpaper,floatfix,10pt]{revtex4-1}
\usepackage{amsmath,amssymb,graphicx,hyperref}
\hypersetup{pdfnewwindow=true, colorlinks=true, linkcolor=blue, anchorcolor=blue,
  citecolor=blue, filecolor=blue, menucolor=blue, urlcolor=blue}

\def\Tc{T_\text{c}}

\begin{document}

\title{Guided design of copper oxysulfide superconductors}

\author{Chuck-Hou Yee}
\email{chuckyee@physics.rutgers.edu}
\author{Turan Birol}
\author{Gabriel Kotliar}
\affiliation{Dept. of Physics \& Astronomy, Rutgers, The State University of New Jersey, Piscataway, NJ 08854, USA}

\date{\today}

\begin{abstract}
  We describe a framework for designing novel materials, combining modern
  first-principles electronic structure tools, materials databases, and
  evolutionary algorithms capable of exploring large configurational spaces.
  Guided by the chemical principles introduced by Antipov, \emph{et. al.}, for
  the design and synthesis of the Hg-based high-temperature superconductors, we
  apply our framework to design a new layered copper oxysulfide,
  Hg(CaS)$_2$CuO$_2$. We evaluate the prospects of superconductivity in this
  oxysulfide using theories based on charge-transfer energies, orbital
  distillation and uniaxial strain.
\end{abstract}


\maketitle

The superconductors with the highest known transition temperatures at ambient
pressure are all layered compounds containing planes of copper and oxygen where
the superconducting electrons reside, separated by ``spacer layers'' composed
of other elements. For a given compound, varying the doping level to an optimal
value near 0.15 holes per copper maximizes the superconducting transition
temperature $\Tc$. Since the copper oxide planes are a common ingredient, the
large variabilty in optimal $\Tc$'s between compounds must then be controlled
by the spacer layers, which function to tune the chemical and structural
properties of the copper oxide layer. Finding novel compositions for the spacer
layers is key to discovering new superconductors with higher transition
temperatures.

Theoretical design of new compounds is challenging due to the vast
combinatorial space of elements and the large number of constraints: preferred
oxidation state, electronegativity, atomic radii, preferred local coordination
environment, and overall electrical neutrality. These microscopic properties in
turn determine the local structural stability, global configurational minimum,
and thermodynamic stability. Finding the global low-energy structure is the
computational bottleneck. If a given composition results in a stable compound,
we still need to determine whether it tunes the low-energy Hamiltonian so that
$\Tc$ is enhanced, which imposes further screening criteria.

Rather than design a compound from scratch, we adopt a more moderate approach:
we take as a starting point the family of cuprates with the highest transition
temperatures, the Hg-based cuprates, and modulate its spacer layers. We benefit
from the vast body of chemical intuition accumulated for the cuprates, for
instance, laid out especially clearly in Ref.~\onlinecite{Antipov2008}, and use
modern electronic structure methods~\cite{kresse1996a,kresse1996b}, materials
databases~\cite{Jain2013, Belsky2002} and evolutionary
algorithms~\cite{Oganov2006} to efficiently screen for compositions which have
the desired structure and the best prospects for stability. We use three
proposed theories, based on epitaxial strain~\cite{Locquet1998}, the
charge-transfer energy~\cite{Weber2012}, and orbital
distillation~\cite{Sakakibara2010}, to evaluate the resultant compound's
prospects for superconducivity. In this manuscript, we describe the general
framework for guided design of new materials, and demonstrate its workflow by
applying the principles to design a new layered copper oxysulfide
Hg(CaS)$_2$CuO$_2$, which we abbreviate (HCSCO). If it can be synthesized and
doped, we believe HCSCO will be a high-temperature superconductor.

\section{Rational Design}

\begin{figure*}
  \includegraphics[width=0.8\textwidth]{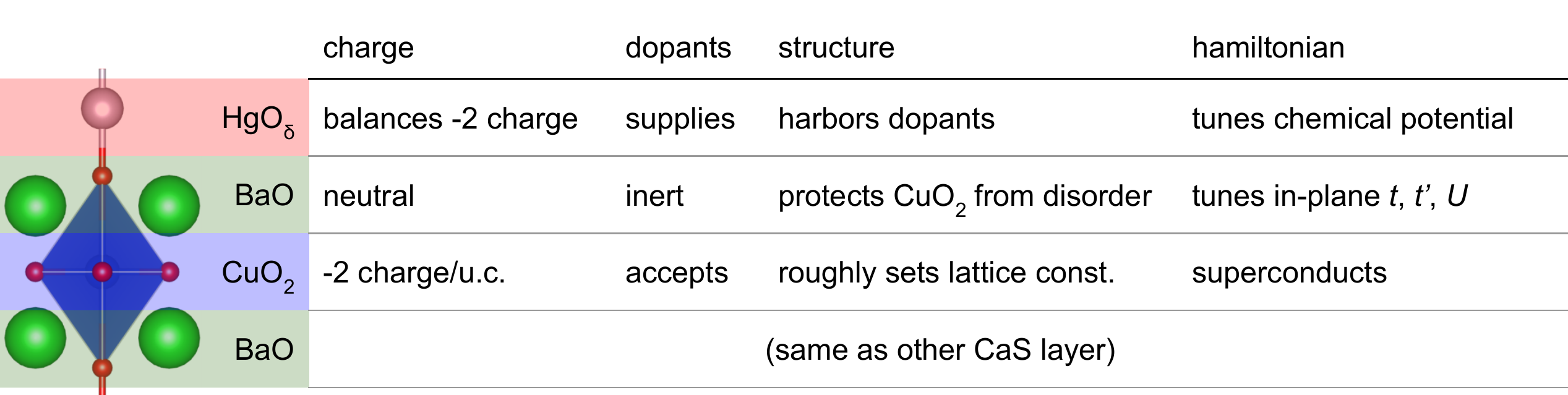}
  \caption{Each layer in the structural stack of HgBa$_2$CuO$_{4+\delta}$
    performs a specific chemical, structural and electronic function.  We
    focused on tuning the layers immediately adjacent to the CuO$_2$ plane
    which function to tune the in-plane hamiltonian, but must also be charge
    neutral and not interfere with doping.}
  \label{fig:functions}
\end{figure*}

The cuprates are a functional stack: the composition of each layer is chosen to
play a specific role (Fig.~\ref{fig:functions}). The central copper oxide
(CuO$_2$) plane supports superconductivity and roughly constrains the in-plane
lattice constant. The remaining layers must tune the chemical potential of the
CuO$_2$ layer without rumpling the plane or introducing disorder, and isolate
each CuO$_2$ plane to create a 2D system.

In the Hg-cuprates, the HgO$_\delta$ layer harbors dopant atoms which tune the
chemical potential. The BaO layers immediately adjacent to the CuO$_2$ plane
spatially separate the superconducting electrons from the detrimental effects
of the disordered dopant layer~\cite{Fujita2005,Hobou2009}. Additionally, the
highly ionic nature of the BaO layer means they do not capture dopant electrons
intended for the CuO$_2$ plane. The preference of Hg to be dumbbell coordinated
bonds the entire structure together without introducing structural
distortions~\cite{Antipov2008}. Finally, the highly ionic O-Hg-O dumbbells
minimize $c$-axis hopping to maintain 2-dimensionality.

Due to their spatial proximity, the adjacent BaO layers tune the hoppings and
interaction strengths of the in-plane Hamiltonian. Designing compounds with
novel adjacent layers provides a mechanism for controlling superconductivity
by, e.g., reducing the charge-transfer energy. We quickly realized the most
stringent constraint is structural stability, so we focused first on isolating
stable candidates, then subsequently investigating their electronic properties.

\begin{figure}
  \includegraphics[width=\columnwidth]{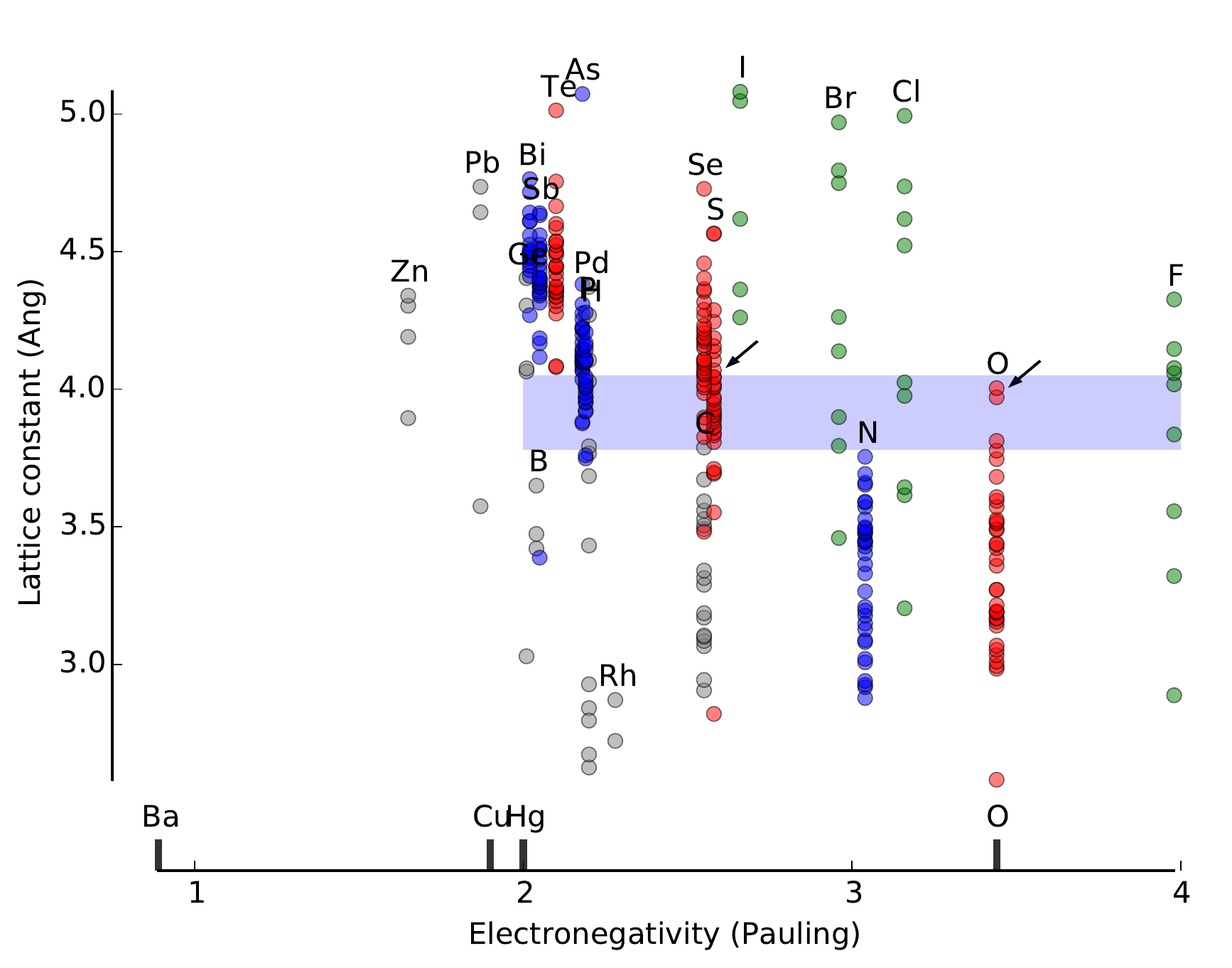}
  \caption{Rock salt compositions AX plotted as a function of their lattice
    constant vs. electronegativity of the anion X, and color-coded by anion's
    nominal valence. Electronegativities of elements existing in
    HgBa$_2$CuO$_4$ are marked. We seek compositions which have (a) lattice
    constants comparable to the in-plane Cu-Cu distance in the cuprates ($\sim$
    3.80-3.95~\AA), and (b) anion electronegativities greater than that of Cu,
    as dopants intended for the copper oxide plane would otherwise be captured
    by the anion, resulting in additional Fermi surfaces. The compositions of
    interest are marked by a shaded rectangle. Arrows point to BaO and CaS.}
  \label{fig:structfield}
\end{figure}

To maximize the likelihood that a proposed composition is stable, we note that
the layers adjacent to the CuO$_2$ plane form a rock salt structure. Using
materials databases, we selected all naturally occurring rock salt compounds
AX, composed of an cation A and an anion X (Fig.~\ref{fig:structfield}). The
phase space is large and the rate limiting step is structural prediction, so we
quickly pre-screen candidates by discarding compositions with (1) large lattice
mismatches relative to the in-plane Cu-Cu distance, which we took to be
3.82~\AA, and (2) anions less electronegative than Cu, as these anions would
capture dopants intended for the superconducting plane, producing additional
Fermi surfaces.

Given the remaining list of potential compositions, we screened for stable
compounds as follows:
\begin{enumerate}
  \item Local stability -- substitute AX for BaO layer of HBCO structure
    forming Hg(AX)$_2$CuO$_2$, fully relax the structure, then compute the
    phonons. Discard if there are unstable phonons.
  \item Configurational minimum -- place one formula unit of the elements (1Hg,
    2A, 2X, 1Cu 2O) in a box and apply evolutionary algorithms to find most
    energetically stable crystal structure.
  \item Thermodynamic stability -- compute convex hull of Hg-A-X-Cu-O system
    and determine whether Hg(AX)$_2$CuO$_2$ lies below the hull. Synthesis
    generally proceeds in oxygen environments so the hull is computed with
    fixed Hg-A-X-Cu stoichiometry and varying oxygen concentration,
    parameterized by the chemical potential $\mu(\text{O}_2)$.
\end{enumerate}
Compositions which pass all these hurdles have a good chance of being
synthesizble in bulk. However, these criteria aren't rigid, as compositions
which exhibit only weak phonon instabilities are likely synthesizable and
possibly superconducting. Additionally, compositions that lie above the convex
hull may also be synthesized via epitaxy or under high pressures.

\section{Methods}

The \texttt{pymatgen} implementation of the Materials API~\cite{Ong2013} was
used to access the Materials Project~\cite{Jain2013} database to select all
structures with lattice constants within $\sim 5$\% of 3.9\AA, the nominal
Cu-Cu distance.

Structural relaxations within DFT were performed with VASP~\cite{kresse1996a,
  kresse1996b} using PAW potentials~\cite{bloechl994} and the PBEsol
functional~\cite{perdew1996}. We computed phonons at the $(0,0)$, $(\pi,0)$ and
$(\pi,\pi)$ $k$-points using $1 \times 2$ and $\sqrt{2} \times \sqrt{2}$
supercells constructed with the help of ISOTROPY~\cite{Stokes2005} to check
local stability.

%% We used an energy convergence criteria of [], a plane wave cutoff of [],
%% k-point grid of [].

We performed a fixed-composition evolutionary search for crystal structure
using USPEX~\cite{Glass2006, Lyakhov2013}, with VASP as the underlying DFT
engine. A unit cell consisting of one formula unit and an initial population of
256 random structures were used, decreasing to 64 in following generations.
USPEX successfully found existing cuprate structures (LSCO, YBCO, HBCO) in
benchmark tests, demonstrating its efficacy in the few-atom (8 atoms/f.u.) but
many-species (5 elements) regime.

The Gibbs phase diagrams for the Hg-A-X-Cu system at varying oxygen potentials
$\mu(\text{O}_2)$ were constructed using the phase diagram implementation in
\texttt{pymatgen}~\cite{PingOng2008}. We cross-checked these results with the
PBE functional, which is known to give improved atomization
energies~\cite{perdew1996}, finding no significant differences in the relative
energies between compounds.

Bandstructure orbital character was analyzed with WIEN2k~\cite{wien2k} and
structures plotted using VESTA~\cite{vesta}. Electronic parameters were
extracted by downfolding to atomic-like orbitals, defined in~\cite{Haule2010}.

\section{Results}

\begin{figure}
  \includegraphics[width=0.8\columnwidth]{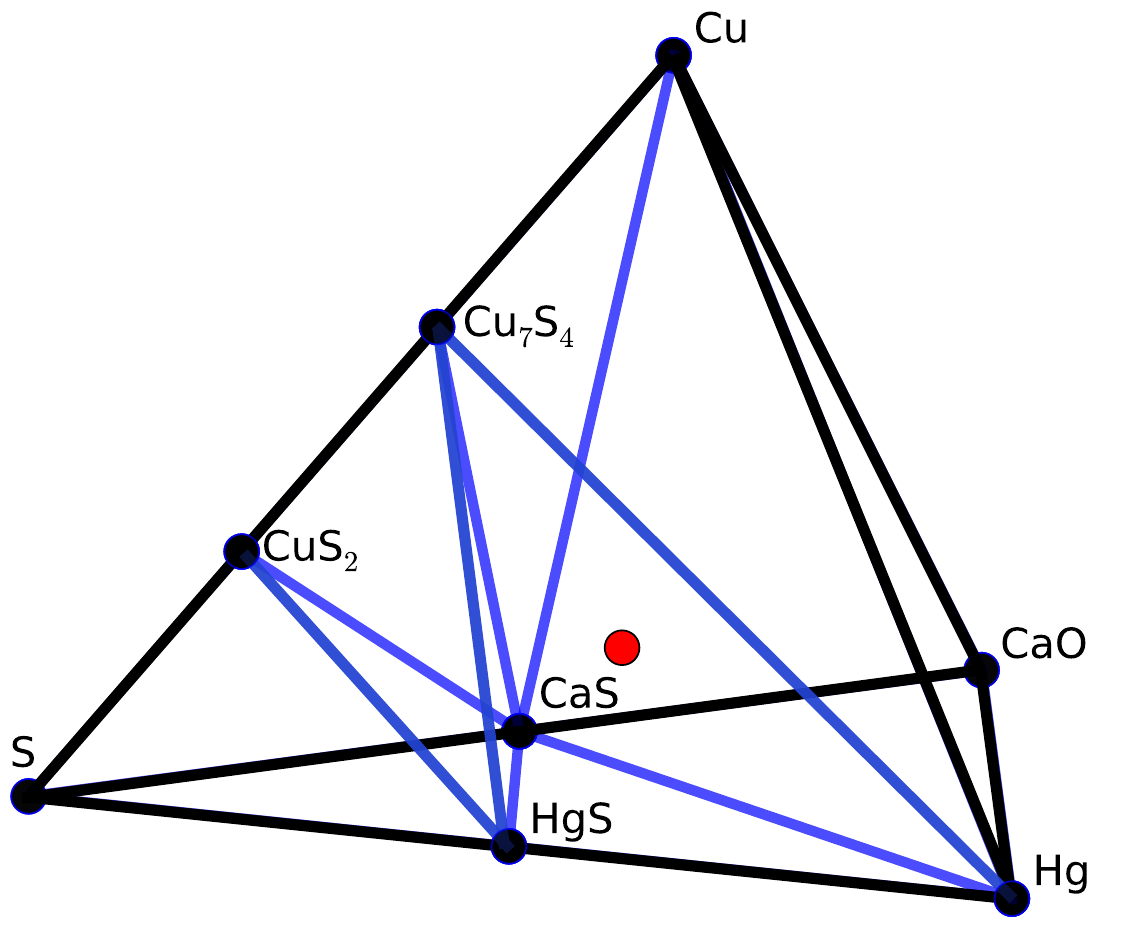}
  \caption{The Gibbs phase diagram of the Hg-Ca-S-Cu system at an oxygen
    chemical potential of $\mu(\text{O}_2) = -16.48~\text{eV}$, chosen because
    Hg(CaS)$_2$CuO$_2$ (red dot) is least unstable at this value. The phase
    diagram forms a tetrahedron with S, Hg, Cu and CaO at the vertices
    (elemental Ca isn't stable under this oxygen environment). HCSCO lies in
    the interior of the tetrahedron, on the triangular face formed by Hg, Cu
    and CaS.}
  \label{fig:gibbs}
\end{figure}

Our materials design framework efficiently isolates compositions with desired
crystal structures and properties. From roughly 350 rock salt compounds,
pre-screening based on lattice constants and electronegativity selected $\sim
20$ compositions. Screening for local stability via phonon modes left three
compositions: CaS, ZrAs, and YbS. Configurational stability using evolutionary
structure prediction showed only a single composition, Hg(CaS)$_2$CuO$_2$,
adopted the HBCO structure. This compound has sulfur instead of oxygen as the
apical atom and has the smaller calcium ion instead barium in the adjacent
layer. We also performed structure prediction with two formula units, finding
the same HBCO structure. Thermodynamic calculations show HCSCO is marginally
unstable, lying 240~meV/atom above the convex hull at $\mu(\text{O}_2) =
-16.48~\text{eV}$, with an expected decomposition into Hg, Cu, CaS, and oxygen
gas (Fig.~\ref{fig:gibbs}).

\begin{figure}
  \includegraphics[width=\columnwidth]{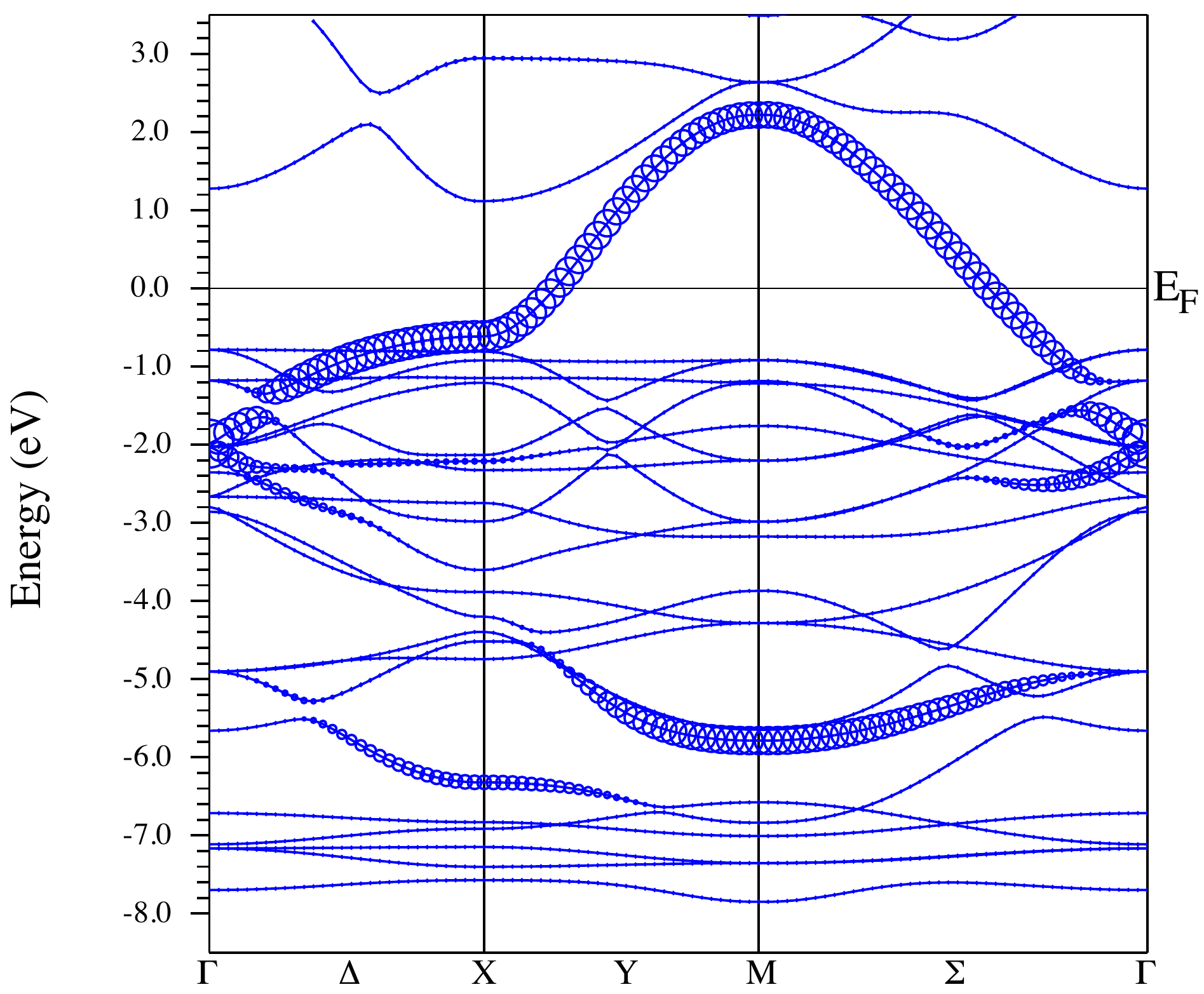}
  \includegraphics[width=\columnwidth]{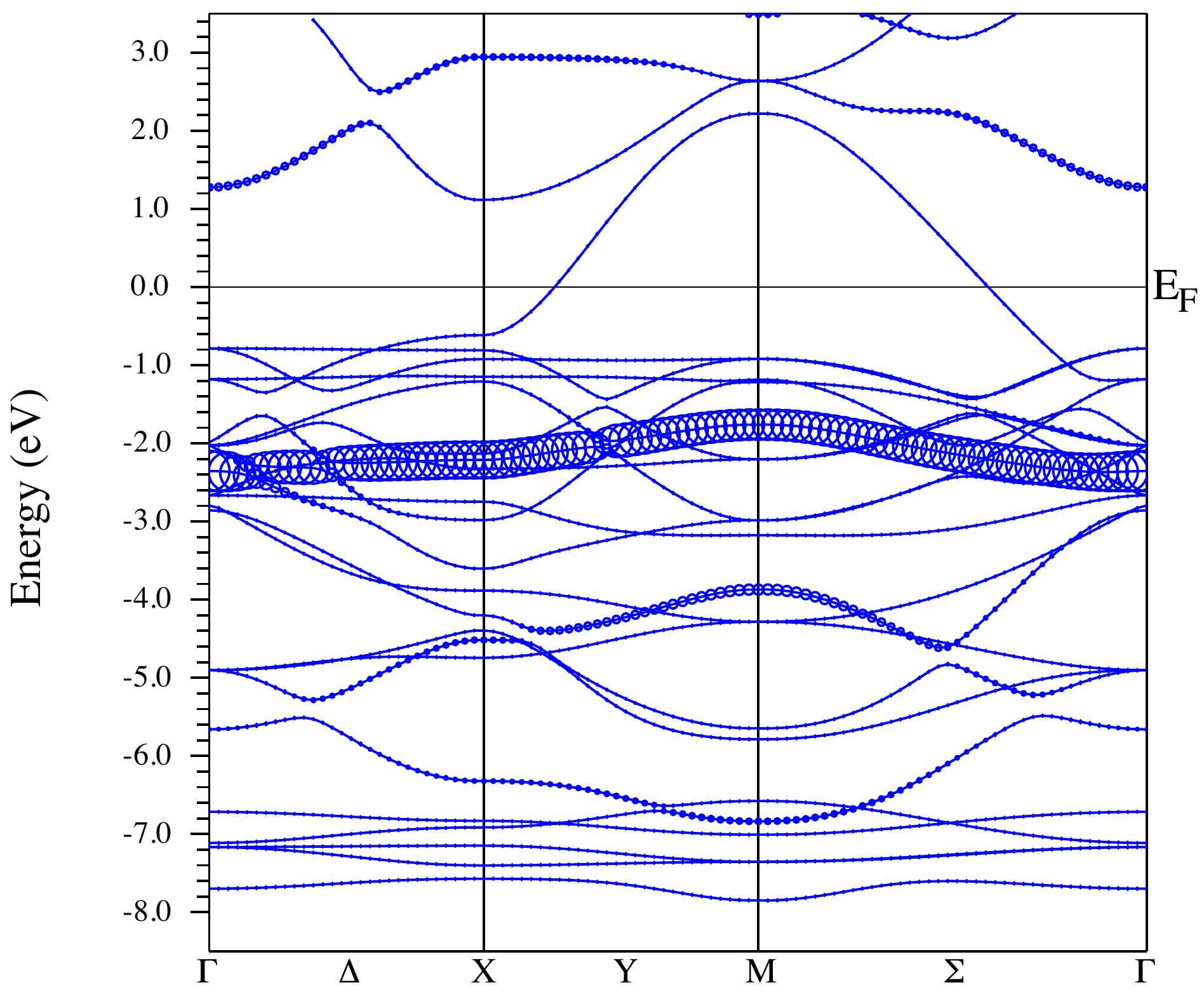}
  \caption{Bandstructure of HCSCO, with $d_{x^2-y^2}$ (top) and $d_{z^2}$
    (bottom) character highlighted. A single band crosses the Fermi level of
    mainly $d_{x^2-y^2}$ character with little admixture of $d_{z^2}$
    character.}
  \label{fig:bands}
\end{figure}

Turning to electronic structure (Fig.~\ref{fig:bands}), we observe a single
band crossing the Fermi level with strong $d_{x^2-y^2}$ character, similar to
other single-layer cuprates. The Fermi surface extremely 2-dimensional, and
comparable to HBCO. Table~\ref{tbl:params} summarizes a comparison of the
salient electronic parameters extracted using downfolding between HCSCO and the
single-layer HBCO superconductor.

\begin{table}[b]
  \begin{tabular}{rccccccccc}
             & \multicolumn{3}{c}{Structure~(\AA)} & & \multicolumn{5}{c}{Parameters~(eV)} \\
    \cline{2-4} \cline{6-10}
             & $a$    & c       & $d_\text{apical}$ & & $\Delta_{pd}$ & $\Delta E_\text{z}$ & $t_{pd}$ & $t_{pp}$ & $t_{pp}'$ \\ \hline
    HCSCO    & 3.833  & 10.054  & 2.746             & & 1.94 & 0.20 & 1.35 & 0.61 & 0.09 \\
    HBCO     & 3.879  &  9.516  & 2.789             & & 1.93 & 1.87 & 1.25 & 0.65 & 0.16 \\
  \end{tabular}
  \caption{Comparison of structural and electronic parameters of
    HgBa$_2$CuO$_4$ and Hg(CaS)$_2$CuO$_2$.}
  \label{tbl:params}
\end{table}

\section{Discussion}

We evaluate the prospects of superconductivity in HCSCO based on three
chemically-based proposals. Applying in-plane compression induces large
increases in transition temperatures, with $d\Tc/d(\log a) \sim -600~\text{K}$
in LSCO~\cite{Locquet1998} and $-4000~\text{K}$ in HBCO~\cite{Antipov2008},
where $a$ is the in-plane lattice constant. We are aware that many variables
are involved, especially that compression may induce buckling, which decreases
$\Tc$~\cite{Gao2009}. However, assuming all other factors are held constant,
the 1.6\% compression of HCSCO relative to HBCO would cause a 60~K increase in
$\Tc$.

\begin{figure}
  \includegraphics[width=\columnwidth]{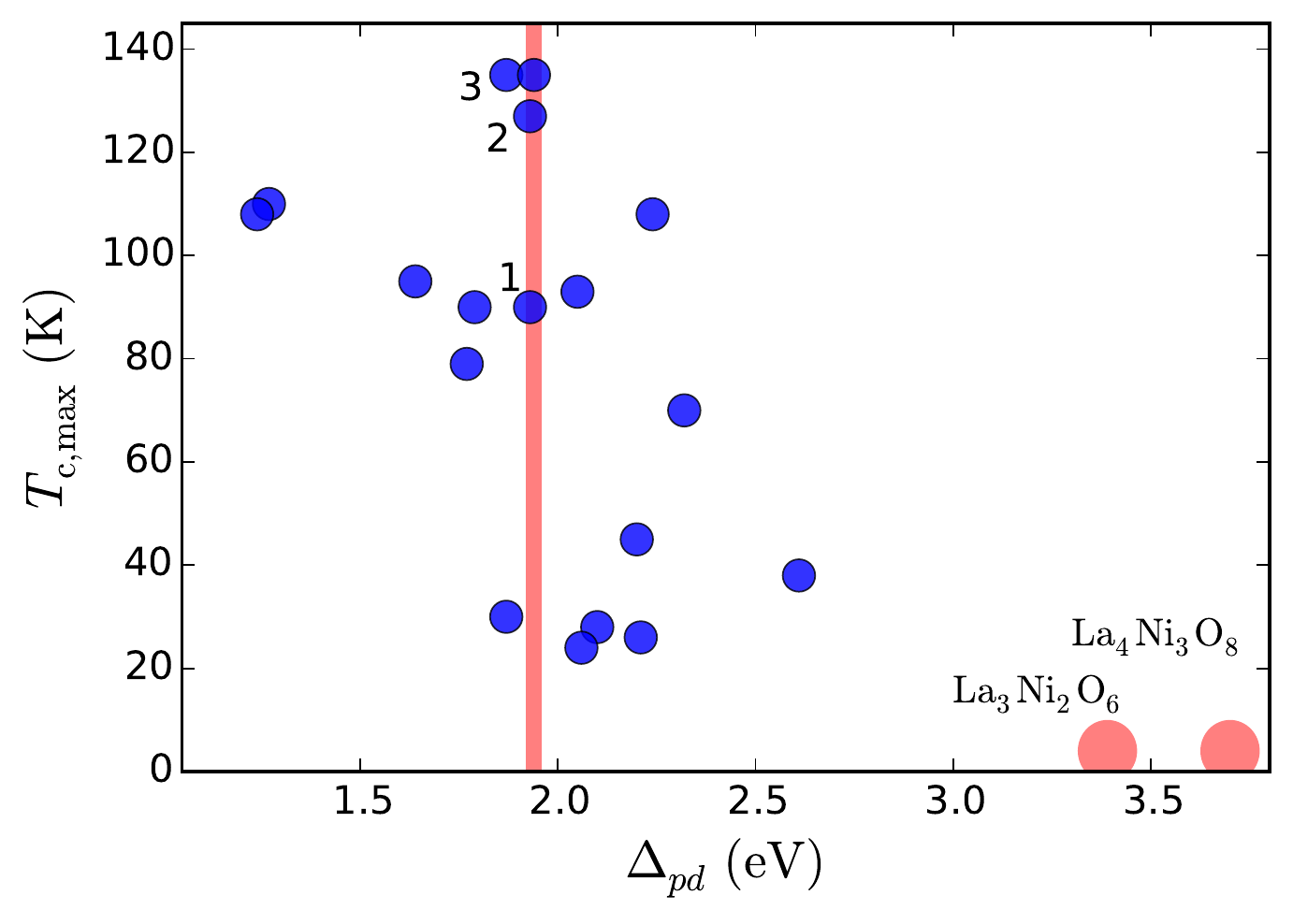}
  \caption{Scatter-plot of the maximal superconducting transition temperature
    $T_\text{c,max}$ vs. charge-transfer energy $\Delta_{pd}$ for the copper
    oxides (see Ref.~\onlinecite{Weber2012} for details). The vertical bar
    marks the $\Delta_{pd}$ of HCSCO, which is close to that of the other
    mercury cuprates (labeled 1, 2 and 3 by the number of CuO$_2$ layers);
    thus, we expect similar transition temperatures. For comparison, we have
    also plotted the charge-transfer energies of two layered nickelates,
    La$_3$Ni$_2$O$_6$~\cite{Poltavets2006} and
    La$_4$Ni$_3$O$_8$~\cite{Poltavets2010}, where the Ni valence can be tuned
    near $3d^9$ and are predicted to be nonsuperconducting (red circles).}
  \label{fig:epd}
\end{figure}

An alternative proposal argues that tuning the charge-transfer energy
$\Delta_{pd}$, which roughly corresponds to the effective Coulomb interaction,
controls superconductivity~\cite{Weber2012}. As shown in
Table~\ref{tbl:params}, the negligible difference in $\Delta_{pd}$ between
HCSCO and HBCO, combined with similar hopping amplitudes, implies comparable
$\Tc$'s (Fig.~\ref{fig:epd}). We expected that the increased polarizability of
sulfur as compared to oxygen would reduce $\Delta_{pd}$, an effect which may be
captured in methods beyond LDA, like the GW approximation.

Finally, the orbital distillation proposal argues that a large admixture of the
apical orbitals, in particular the Cu-$d_{z^2}$ orbital, into the $d_{x^2-y^2}$
band suppresses $\Tc$~\cite{Sakakibara2010}. Although the energy splitting
$\Delta E_z \equiv E_{x^2-y^2} - E_{z^2}$ is quite small
(Table~\ref{tbl:params}), there is no admixture of apical orbital charcter into
the in-plane band (Fig.~\ref{fig:bands}). Orbital distillation predicts $\Tc$'s
comparable to the HBCO superconductors.

Turning to stability against phase separation, our calculations indicate HCSCO
lies slightly above the convex hull, well within the systematic errors of the
theory used to estimate total energies. Furthermore it is known that many
interesting functional materials are metastable~\cite{Zhang2012}, protected
from decay by large energetic barriers. In solid state synthesis, the
Hg-cuprates are formed in the presence of gaseous Hg~\cite{Alyoshin2002} in
addition to O$_2$, providing an additional tuning parameter which may enhance
stability. Finally, prior experience with materials design indicates that
compounds which are too stable may not be easily dopable~\cite{Yin2013,
  Retuerto2015}. These considerations suggest avenues for synthesis of HCSCO
should be explored, both under pressure in bulk and especially via molecular
beam epitaxy.

\section{Conclusions}

We have created a workflow for systematically designing novel compositions
based on existing compound families. Guided by chemical intuition and materials
databases, our workflow quickly isolates promising compositions for in-depth
examination. Combining electronic structure tools, evolutionary algorithms, and
methods for constructing phase diagram, we screen for local, configurational
and thermodynamic stability to isolate new compounds with a high probability of
synthesis.

We have applied this workflow to design a new layered copper oxysulfide,
Hg(CaS)$_2$CuO$_2$, which we believe will be a high-temperature superconductor,
with $\Tc$'s comparable to those of the mercury cuprates. In the future, we
expect the multi-layer family members of HCSCO will be designed, with formulas
Hg(CaS)$_2$R$_{n-1}$(CuO$_2$)$_n$ where R is a 2+ cation placed between the
CuO$_2$ planes in the ``infinite-layer'' stack, as well as systematic
extensions to the compositions of other cuprate families.

To date, cuprate superconductors exist with oxygen, chlorine and fluorine in
the apical position. The synthesis HCSCO would produce the exciting prospect of
adding a fourth, sulfur-based family.

%% Additionally, we expect structural distortions and chemical this order to be
%% minimized because we designed this compound in accordance to the principles set
%% forth by Antipov.

\section{Acknowledgements}
C.Y. and G.K. were supported by the DOE EFRC CES program. T.B. acknowledges the
support of the Center for Materials Theory at Rutgers University.

\bibliography{hcsco,vaspref,otherrefs}

\end{document}